# Spinless particles with unequal Scalar-Vector Yukawa interactions

M. Hamzavi[1*], S. M. Ikhdair[2**]

[1]*Department of Basic Sciences, Shahrood Branch, Islamic Azad University, Shahrood, Iran*

[2]*Physics Department, Near East University, 922022 Nicosia, North Cyprus, Mersin 10, Turkey*

[*]*Corresponding author: Tel.:+98 273 3395270, fax: +98 273 3395270*

[*]Email: majid.hamzavi@gmail.com

[**]Email: sikhdair@neu.edu.tr

## Abstract

We present analytical solutions of the spin-zero Klein-Gordon (KG) bosons in the field of unequal mixture of scalar and vector Yukawa potential within the framework of the approximation to the centrifugal potential term for any arbitrary $l$-state. The explicit forms of the energy bound states including energy spectra and unnormalized wave functions are obtained using a simple shortcut of the Nikiforov-Uvarov (NU) method. Our numerical tests using energy calculations demonstrate the existence of inter dimensional degeneracy amongst energy states of the present quantum system consisting of the KG-Yukawa problem. The dependence of the energy on the dimension $D$ is numerically discussed for $D=3-10$.

**Keywords**: Klein-Gordon equation, Yukawa Potential, $D$-dimensional space, Nikiforov-Uvarov Method

**PACS**: 03.65.Fd, 03.65.pm, 03.65.ca, 03.65.Ge.

## 1. Introduction

The Yukawa potential is often used to compute bound-state normalizations and energy levels of neutral atoms [1-6] which have been studied over the past years. It is known that SSCP yields reasonable results only for the innermost states when $Z$ is large. However, for the outermost and middle atomic states, it gives rather poor results. Bound-state energies for the SSCP with $Z=1$, have been studied in the light of the shifted large-$N$ method [7]. Chakrabarti and Das presented a perturbative solution of the Riccati equation leading to analytical superpotential for Yukawa potential [8]. Ikhdair and Sever investigated energy levels of neutral atoms by applying an alternative perturbative scheme in solving the Schrödinger equation for the Yukawa potential model with a modified screening parameter [9]. They also studied bound states of the Hellmann potential, which represents the superposition of the attractive Coulomb potential $-a/r$ and the Yukawa potential $b\exp(-\delta r)/r$ of arbitrary strength $b$ and screening parameter $\delta$ [10].

The aim of the present work is to investigate the KG equation in arbitrary dimension $D$ [11] with unequal mixture of scalar and vector Yukawa potentials:

$$V\left(r\right)=-V_0\frac{e^{-ar}}{r}, \tag{1a}$$

$$S(r)=-S_0\frac{e^{-ar}}{r} \tag{1b}$$

$$S(r)=\beta V(r), \quad -1\leq\beta\leq 1 \tag{1c}$$

where $V_0=\alpha Z$, $\alpha=(137.037)^{-1}$ is the fine-structure constant and $Z$ is the atomic number and $a$ is the screening parameter [1]. Also, $\beta$ is an arbitrary a constant demonstrating the ratio of scalar potential to the vector potential [12]. When $\beta=1$, we have equal mixture, i.e., $S(r)=V(r)$, representing the exact spin symmetric limit $\Delta(r)=S(r)-V(r)=0$ (the potential difference is exactly zero). On the other hand, when $\beta=-1$, we have $S(r)=-V(r)$, representing the exact pseudospin symmetric limit $\Sigma(r)=S(r)+V(r)=0$ (the potential sum is exactly zero). The strong singular centrifugal term is being approximated within the framework of an improved approximation scheme. Further, the spinless $D$-dimensional KG equation is solved using the parametric generalization of the NU method [13,14] to obtain energy eigenvalues and corresponding wave functions for any $l$-state.

The present work is arranged as follows. In section 2, the generalized parametric NU method is briefly introduced. In section 3, we give an introduction to KG equation in $D$-dimensional space and then we obtain the bound state solution of the hyperradial KG equation with Yukawa potential by using NU method. The numerical results are also given. Finally, we end with our concluding remarks in section 4.

## 3. Parametric NU Method

This powerful mathematical tool solves second order differential equations. Let us consider the following differential equation [13,14]

$$\psi_n''(s)+\frac{\tilde{\tau}(s)}{\sigma(s)}\psi_n'(s)+\frac{\tilde{\sigma}(s)}{\sigma^2(s)}\psi_n(s)=0\,, \tag{2}$$

where $\sigma(s)$ and $\tilde{\sigma}(s)$ are polynomials, at most of second degree, and $\tilde{\tau}(s)$ is a first-degree polynomial. To make the application of the NU method simpler and direct without need to check the validity of solution. We present a shortcut for the method. Hence, firstly we write the general form of the Schrödinger-like equation (2) in a more general form as [14]

$$\psi_n''(s)+\left(\frac{c_1-c_2 s}{s(1-c_3 s)}\right)\psi_n'(s)+\left(\frac{-p_2 s^2+p_1 s-p_0}{s^2(1-c_3 s)^2}\right)\psi_n(s)=0, \tag{3}$$

satisfying the wave functions

$$\psi_n(s)=\phi(s)y_n(s). \tag{4}$$

Secondly, we compare (3) with its counterpart (2) to obtain the following parameter values,

$$\tilde{\tau}(s)=c_1-c_2 s,\quad \sigma(s)=s(1-c_3 s),\quad \tilde{\sigma}(s)=-p_2 s^2+p_1 s-p_0, \tag{5}$$

Now, following the NU method [13], we obtain the energy equation [14]

$$c_2 n-(2n+1)c_5+(2n+1)\left(\sqrt{c_9}-c_3\sqrt{c_8}\right)+n(n-1)c_3+c_7+2c_3c_8-2\sqrt{c_8c_9}=0, \tag{6}$$

and the corresponding wave functions

$$\rho(s)=s^{c_{10}}(1-c_3 s)^{c_{11}}\,,\quad \phi(s)=s^{c_{12}}(1-c_3 s)^{c_{13}}\,,\ c_{12}>0,\ c_{13}>0,$$

$$y_n(s)=P_n^{(c_{10},c_{11})}(1-2c_3 s),\ c_{10}>-1,\ c_{11}>-1,$$

$$\psi_{n\kappa}(s)=N_{n\kappa}s^{c_{12}}(1-c_3 s)^{c_{13}}P_n^{(c_{10},c_{11})}(1-2c_3 s). \tag{7}$$

where $P_n^{(\mu,\nu)}(x),\ \mu>-1,\ \nu>-1$, and $x\in[-1,1]$ are Jacobi polynomials with the following constants:

$$c_4=\frac{1}{2}(1-c_1),\qquad c_5=\frac{1}{2}(c_2-2c_3),$$

$$c_6=c_5^2+p_2,\qquad c_7=2c_4c_5-p_1,$$

$$c_8 = c_4^2 + p_0, \qquad c_9 = c_3\left(c_7 + c_3 c_8\right) + c_6,$$

$$c_{10} = c_1 + 2c_4 - 2\sqrt{c_8} - 1 > -1, \qquad c_{11} = 1 - c_1 - 2c_4 + \frac{2}{c_3}\sqrt{c_9} > -1,\ c_3 \neq 0,$$

$$c_{12} = c_4 - \sqrt{c_8} > 0, \qquad c_{13} = -c_4 + \frac{1}{c_3}(\sqrt{c_9} - c_5) > 0,\ c_3 \neq 0, \tag{8}$$

where $c_{12} > 0,\ c_{13} > 0$ and $s \in [0, 1/c_3],\ c_3 \neq 0.$

In the special case when $c_3 = 0$, the wave function (4) becomes

$$\lim_{c_3 \to 0} P_n^{(c_{10}, c_{11})}(1 - 2c_3 s) = L_n^{c_{10}}(c_{11} s), \quad \lim_{c_3 \to 0}(1 - c_3 s)^{c_{13}} = e^{c_{13} s},$$

$$\psi(s) = N s^{c_{12}} e^{c_{13} s} L_n^{c_{10}}(c_{11} s). \tag{9}$$

## 3. Hyperradial part of the KG equation in $D$ -dimensional space

In spherical coordinates, the KG equation with vector $V(r)$ and scalar $S(r)$ potentials can be written as $\left(\text{in units of } \hbar = c = 1\right)$

$$\left[\Delta_D + \left(E_{nl} - V(r)\right)^2 - \left(M + S(r)\right)^2\right]\psi_{nlm}(r, \Omega_D) = 0, \tag{10}$$

with

$$\Delta_D = \nabla_D^2 = \frac{1}{r^{D-1}}\frac{\partial}{\partial r}\left(r^{D-1}\frac{\partial}{\partial r}\right) - \frac{\Lambda_D^2(\Omega_D)}{r^2}, \tag{11}$$

where $E_{nl}$, $\Lambda_D^2(\Omega_D)/r^2$ and $\Omega_D$ are the energy eigenvalues, generation of the centrifugal barrier for $D$-dimensional space and the angular coordinates, respectively [11]. The eigenvalues of the $\Lambda_D^2(\Omega_D)$ are given by [11]

$$\Lambda_D^2(\Omega_D) Y_l^m(\Omega_D) = \frac{\left(D + 2l - 2\right)^2 - 1}{4} Y_l^m(\Omega_D), \quad D > 1 \tag{12}$$

where $Y_l^m(\Omega_D)$ is the hyperspherical harmonics. For $D = 3,$ we have a familiar form of (12) as $\Lambda_{D=3}^2(\Omega_{D=3}) Y_l^m(\Omega_{D=3}) = l(l+1) Y_l^m(\Omega_{D=3})$. Using the procedure of separation of variables by means of the wave function $\psi_{nlm}(r, \Omega_D) = r^{-(D-1)/2} R_{nl}(r) Y_l^m(\Omega_D)$, Eq. (10) reduces to

$$\left[\frac{d^2}{dr^2}-\left(M^2-E_{nl}{}^2\right)-2\left(E_{nl}V(r)+MS(r)\right)+V^2(r)-S^2(r)-\frac{\left(D+2l-2\right)^2-1}{4r^2}\right]R_{nl}(r)=0, \qquad (13)$$

Substituting Eq. (1) into Eq. (13), one obtains

$$\left[\frac{d^2}{dr^2}-\varepsilon^2+\frac{(V_0^2-S_0^2)}{r^2}e^{-2ar}+\frac{2(MS_0+E_{nl}V_0)}{r}e^{-ar}-\frac{\left(D+2l-2\right)^2-1}{4r^2}\right]R_{nl}(r)=0, \qquad (14)$$

where $\varepsilon^2=M^2-E_{nl}^2$. It is obvious that the above equation does not admit an exact solution due to the presence of the singular terms $1/r$ and $1/r^2$. So we resort to approximation schemes to make approximations for these two terms [15-19] as

$$\frac{1}{r^2}\approx 4a^2\frac{e^{-2ar}}{(1-e^{-2ar})^2}, \qquad (15a)$$

$$\frac{1}{r}\approx 2a\frac{e^{-ar}}{(1-e^{-2ar})}, \qquad (15b)$$

which are valid for $ar \ll 1$ [15]. Thus, the Yukawa potential in (1a) and (1b) can be approximated as

$$V(r)=-2aV_0\frac{e^{-2ar}}{\left(1-e^{-2ar}\right)}, \qquad (16a)$$

$$S(r)=-2aS_0\frac{e^{-2ar}}{\left(1-e^{-2ar}\right)}. \qquad (16b)$$

To see the accuracy of our approximation, we plot the Yukawa potential (1a) and its approximation (16a) with parameter values $V_0=\sqrt{2}$, $a=0.05V_0$ [20] as shown in Figure 1. Now, substituting (15) into (13), we obtain

$$\left\{\frac{d^2}{dr^2}-\varepsilon^2+4a^2(V_0^2-S_0^2)\frac{e^{-4ar}}{(1-e^{-2ar})^2}\right.$$
$$\left.+4a(MS_0+E_{n,l}V_0)\frac{e^{-2ar}}{(1-e^{-2ar})}-a^2\left(D+2l-1\right)\left(D+2l-3\right)\frac{e^{-2ar}}{(1-e^{-2ar})^2}\right\}R_{nl}(r)=0, \qquad (17)$$

Making a suitable change of variables as $s=e^{-2ar}$, we can thereby recast Eq. (17) as follows

$$\frac{d^2R_{n,l}(s)}{ds^2}+\frac{1-s}{s(1-s)}\frac{dR_{n,l}(s)}{ds}+\frac{1}{s^2(1-s)^2}\left[-\frac{\varepsilon^2}{4a^2}(1-s)^2+(V_0^2-S_0^2)s^2\right.$$

$$\left.+\frac{(MS_0+E_{nl}V_0)}{a}s(1-s)-\frac{(D+2l-2)^2-1}{4}s\right]R_{nl}(s)=0. \qquad (18)$$

The solution of Eq. (18) can be found by comparing it with Eq. (3) to get

$$c_1=1, \qquad p_0=\frac{\varepsilon^2}{4a^2},$$

$$c_2=1, \qquad p_1=\frac{2\varepsilon^2}{4a^2}+\frac{(MS_0+E_{nl}V_0)}{a}-\frac{(D+2l-2)^2-1}{4}$$

$$c_3=1, \qquad p_2=\frac{\varepsilon^2}{4a^2}-(V_0^2-S_0^2)+\frac{(MS_0+E_{nl}V_0)}{a}. \qquad (19)$$

The values of coefficients $c_i$ $(i=4,5,...,13)$ are found from Eq. (8) and displayed in table 1. Thus, using Eq. (6), the energy eigenvalue equation can be obtained as follows

$$\left(2n+1+\sqrt{(D+2l-2)^2-4(V_0^2-S_0^2)}-\frac{1}{a}\sqrt{M^2-E_{nl}^2}\right)^2$$

$$=\frac{M^2-E_{nl}^2}{a^2}+\frac{4(MS_0+E_{nl}V_0)}{a}+4\left(S_0^2-V_0^2\right), \qquad (20a)$$

or it can be simply expressed as

$$\left(2n+1+\sqrt{(D+2l-2)^2+4\left(S_0^2-V_0^2\right)}-\frac{1}{a}\sqrt{M^2-E_{nl}^2}\right)^2=-\left(\frac{E_{nl}}{a}-2V_0\right)^2+\left(\frac{M}{a}+2S_0\right)^2, \qquad (20b)$$

where $-M<E_{nl}<M$. Some numerical results are given in Tables 2 to 4. The potential parameters are taken as $a=0.05\ fm$ and $M=1.0\ fm^{-1}$ [18]. In Table 2, we calculate the energy states for various values of dimensions $D$ and quantum numbers $n$ and $l$ when vector potential is different from scalar one, i.e. $V(r)\neq S(r)$. In Table 3, we obtain the energy levels for various values of dimensions $D$ and quantum numbers $n$ and $l$ when vector potential is equal to scalar one, i.e. $V(r)=S(r)$. Finally, in Table 4, we obtain the bound state energy for various values of dimensions $D$ and quantum numbers $n$ and $l$ when vector potential is opposite to scalar one, i.e. $V(r)=-S(r)$. As we can see from these Tables, there is an inter-dimensional degeneracy of eigenvalues of upper/lower dimensional system (shown in boldface) from the well-known eigenvalues of a lower/upper dimensional system by means of the transformation $(n,l,D)\to(n,l\pm1,D\mp2)$. Further, it is noticed that the energy of the state becomes less attractive with the increasing of the quantum numbers $n$, $l$ and the dimensional space $D$. In the case when $S_0=\pm V_0$ (i.e., $\beta=\pm1$), the energy levels: $E_{21}=E_{30}$, for any arbitrary dimension $D=3-10$. However, when $\beta=0.5$ the energy levels $E_{21}\neq E_{30}$, that is, the degeneracy is removed.

On the other hand, in the calculations of the corresponding eigenfunctions, we use the relations in (7) to obtain

$$\rho(s)=s^{\frac{\varepsilon}{a}}\left(1-s\right)^{\sqrt{4(S_0^2-V_0^2)+(D+2l-2)^2}},$$

$$\phi(s)=s^{-\frac{\varepsilon}{2a}}\left(1-s\right)^{\frac{1}{2}\left(1+\sqrt{4(S_0^2-V_0^2)+(D+2l-2)^2}\right)},$$

$$y_n(s)=P_n^{\left(\frac{\varepsilon}{a},\sqrt{4(S_0^2-V_0^2)+(D+2l-2)^2}\right)}\left(1-2s\right),$$

$$R_{nl}(s)=N_{nl}s^{-\frac{\varepsilon}{2a}}\left(1-s\right)^{\frac{1}{2}\left(1+\sqrt{4(S_0^2-V_0^2)+(D+2l-2)^2}\right)}P_n^{\left(\frac{\varepsilon}{a},\sqrt{4(S_0^2-V_0^2)+(D+2l-2)^2}\right)}\left(1-2s\right), \tag{21}$$

where $N_{nl}$ is the normalization constant. It can also be rewritten in a more convenient form in terms of the potential parameters as

$$R_{nl}(r)=N_{nl}e^{-\varepsilon r}\left(1-e^{-2ar}\right)^{\frac{1}{2}\left(1+\sqrt{4(S_0^2-V_0^2)+(D+2l-2)^2}\right)}$$
$$\times P_n^{\left(\frac{\varepsilon}{a},\sqrt{4(S_0^2-V_0^2)+(D+2l-2)^2}\right)}\left(1-2e^{-2ar}\right). \tag{22}$$

### 3.1. Some special cases

When we set $V(r)\to V(r)/2$, $S(r)\to S(r)/2$, $E_{nl}+M\to 2\mu/\hbar^2$ and $E_{nl}-M\to E_{nl}$ [17], we obtain the solution in the non-relativistic limit of the Yukawa problem. Here $\mu=m_1m_2/(m_1+m_2)$ is the reduced mass with $m_1$ and $m_2$ stand for masses of the electron $e$ and the atom $Ze$, respectively. Under these conditions, one can obtain the non-relativistic energy eigenvalues of the Yukawa potential as [19]

$$E_{nl} = -\frac{\hbar^2}{2\mu}\left[\left(n+\frac{D}{2}+l-\frac{1}{2}\right)a - \frac{\mu V_0}{\hbar^2\left(n+\frac{D}{2}+l-\frac{1}{2}\right)}\right]^2, \tag{23}$$

and the corresponding radial wave functions reduce to

$$R_{nl}(r) = N_{n,l}\, e^{-\frac{1}{\hbar}\sqrt{2\mu E_{nl}}\, r}(1-e^{-2ar})^{l+1} P_n^{\left(\frac{1}{\hbar a}\sqrt{2\mu E_{nl}},\, 2l+1-\frac{1}{\hbar a}\sqrt{2\mu E_{nl}}\right)}(1-2e^{-2ar}). \tag{24}$$

Also, when the screening parameter $a$ approaches zero, the potential (1) reduces to a Coulomb potential. Thus, in this limit the energy eigenvalues of (23) become the energy levels of the Coulomb interaction, i.e.

$$E_{nl} = -\frac{\mu V_0^2}{2\hbar^2\left(n+\frac{D}{2}+l-\frac{1}{2}\right)^2}, \tag{25}$$

which is identical to Refs. [11, 19] when $D=3$.

## 4. Concluding Remarks

We have used a simple shortcut of the NU method as well as an appropriate approximation for the strong and soft singular terms to obtain approximate analytical bound states solutions of the $D$-dimensional KG equation for scalar and vector Yukawa potentials. Numerical tests using energy calculations show the existence of inter-dimensional degeneracy of energy states prevailing to the following transformation: $(n,l,D)\rightarrow(n,l\pm 1,D\mp 2)$ as shown in Tables 2 to 4. In the present solution, we have not faced any cumbersome and time-consuming procedures in obtaining the numerical and analytical eigenvalues and wave functions of the problem by a simple methodology. The results are applicable to a wide range of relevant fields.

**Table 1.** The values of the parametric constants used to calculate the energy eigenvalues and eigenfunctions

| constant | Analytical value |
|---|---|
| $c_4$ | $0$ |
| $c_5$ | $-\frac{1}{2}$ |
| $c_6$ | $\frac{1}{4}+\frac{\varepsilon^2}{4a^2}-(V_0^2+S_0^2)+\frac{(MS_0+E_{n,l}V_0)}{a}$ |
| $c_7$ | $-\frac{2\varepsilon^2}{4a^2}-\frac{(MS_0+E_{n,l}V_0)}{a}+\frac{(D+2l-1)(D+2l-3)}{4}$ |
| $c_8$ | $\frac{\varepsilon^2}{4a^2}$ |
| $c_9$ | $-(V_0^2+S_0^2)+\frac{(D+2l-1)(D+2l-3)+1}{4}$ |
| $c_{10}$ | $2\sqrt{\frac{\varepsilon^2}{4a^2}}$ |
| $c_{11}$ | $2\sqrt{-(V_0^2+S_0^2)+\frac{(D+2l-1)(D+2l-3)+1}{4}}$ |
| $c_{12}$ | $-\sqrt{\frac{\varepsilon^2}{4a^2}}$ |
| $c_{13}$ | $\frac{1}{2}+\sqrt{-(V_0^2+S_0^2)+\frac{(D+2l-1)(D+2l-3)+1}{4}}$ |

**Table 2.** The energy levels of the Yukawa potential for various $D$ , $n$ and $l$ values with $V_0 = 0.2$ and $S_0 = 0.1$, where $(\beta = 0.5)$.

| | $E_{nl}(fm^{-1})$ | | | | | |
|---|---|---|---|---|---|---|
| $n,l$ D | 1,0 | 2,0 | 2,1 | 3,0 | 3,1 | 3,2 |
| **3** | -0.98885705 | -0.98338741 | **-0.97466557** | -0.97487357 | -0.96331623 | **-0.94911313** |
| **4** | -0.98646010 | -0.97934739 | **-0.96930758** | -0.96939381 | -0.95656399 | **-0.94094526** |
| **5** | -0.98323850 | **-0.97466557** | **-0.96326494** | -0.96331623 | **-0.94911313** | **-0.93204148** |
| **6** | -0.97928221 | **-0.96930758** | **-0.95652848** | -0.95656399 | **-0.94094526** | **-0.92238088** |
| **7** | -0.97462541 | **-0.96326494** | **-0.94908636** | **-0.94911313** | **-0.93204148** | **-0.91193997** |
| **8** | -0.96927908 | **-0.95652848** | **-0.94092392** | **-0.94094526** | **-0.92238088** | **-0.90069216** |
| **9** | -0.96324303 | **-0.94908636** | -0.93202380 | **-0.93204148** | **-0.91193997** | -0.88860737 |
| **10** | -0.95651076 | **-0.94092392** | -0.92236581 | **-0.92238088** | **-0.90069216** | -0.87565154 |

**Table 3.** Bound state energy levels of the Yukawa potential for various $D$, $n$ and $l$ values with $V_0 = S_0 = 0.2$.

| | $E_{n,l}(fm^{-1})$ | | | | | |
|---|---|---|---|---|---|---|
| $n,l$ <br> D | 1,0 | 2,0 | 2,1 | 3,0 | 3,1 | 3,2 |
| **3** | -0.99503719 | -0.98879900 | **-0.97999949** | -0.97999949 | -0.96856958 | **-0.95441573** |
| **4** | -0.99223481 | -0.98472320 | **-0.97461857** | -0.97461857 | -0.96184005 | **-0.94628043** |
| **5** | -0.98879900 | **-0.97999949** | **-0.96856958** | -0.96856958 | **-0.95441573** | **-0.93741586** |
| **6** | -0.98472320 | **-0.97461857** | **-0.96184005** | -0.96184005 | **-0.94628043** | **-0.92780131** |
| **7** | -0.97999949 | **-0.96856958** | **-0.95441573** | **-0.95441573** | **-0.93741586** | **-0.91741347** |
| **8** | -0.97461857 | **-0.96184005** | **-0.94628043** | **-0.94628043** | **-0.92780131** | **-0.90622603** |
| **9** | -0.96856958 | **-0.95441573** | -0.93741586 | **-0.93741586** | **-0.91741347** | -0.89420931 |
| **10** | -0.96184005 | **-0.94628043** | -0.92780131 | **-0.92780131** | **-0.90622603** | -0.88132977 |

**Table 4.** Bound state energy levels of the Yukawa potential for various $D$ , $n$ and $l$ values with $V_0 = 0.2$ and $S_0 = -0.2$ .

| | $E_{nl}\,(fm^{-1})$ | | | | | |
|---|---|---|---|---|---|---|
| $n,l$ D | 1,0 | 2,0 | 2,1 | 3,0 | 3,1 | 3,2 |
| **3** | -0.95533246 | -0.95980903 | **-0.95464935** | -0.95464935 | -0.94475060 | **-0.93125228** |
| **4** | -0.95948526 | -0.95796541 | **-0.95018875** | -0.95018875 | -0.93842313 | **-0.92325937** |
| **5** | -0.95980903 | **-0.95464935** | **-0.94475060** | -0.94475060 | **-0.93125228** | **-0.91445014** |
| **6** | -0.95796541 | **-0.95018875** | **-0.93842313** | -0.93842313 | **-0.92325937** | **-0.90481957** |
| **7** | -0.95464935 | **-0.94475060** | **-0.93125228** | **-0.93125228** | **-0.91445014** | **-0.89435454** |
| **8** | -0.95018875 | **-0.93842313** | **-0.92325937** | **-0.92325937** | **-0.90481957** | **-0.88303523** |
| **9** | -0.94475060 | **-0.93125228** | -0.91445014 | **-0.91445014** | **-0.89435454** | -0.87083573 |
| **10** | -0.93842313 | **-0.92325937** | -0.90481957 | **-0.90481957** | **-0.88303523** | -0.85772427 |

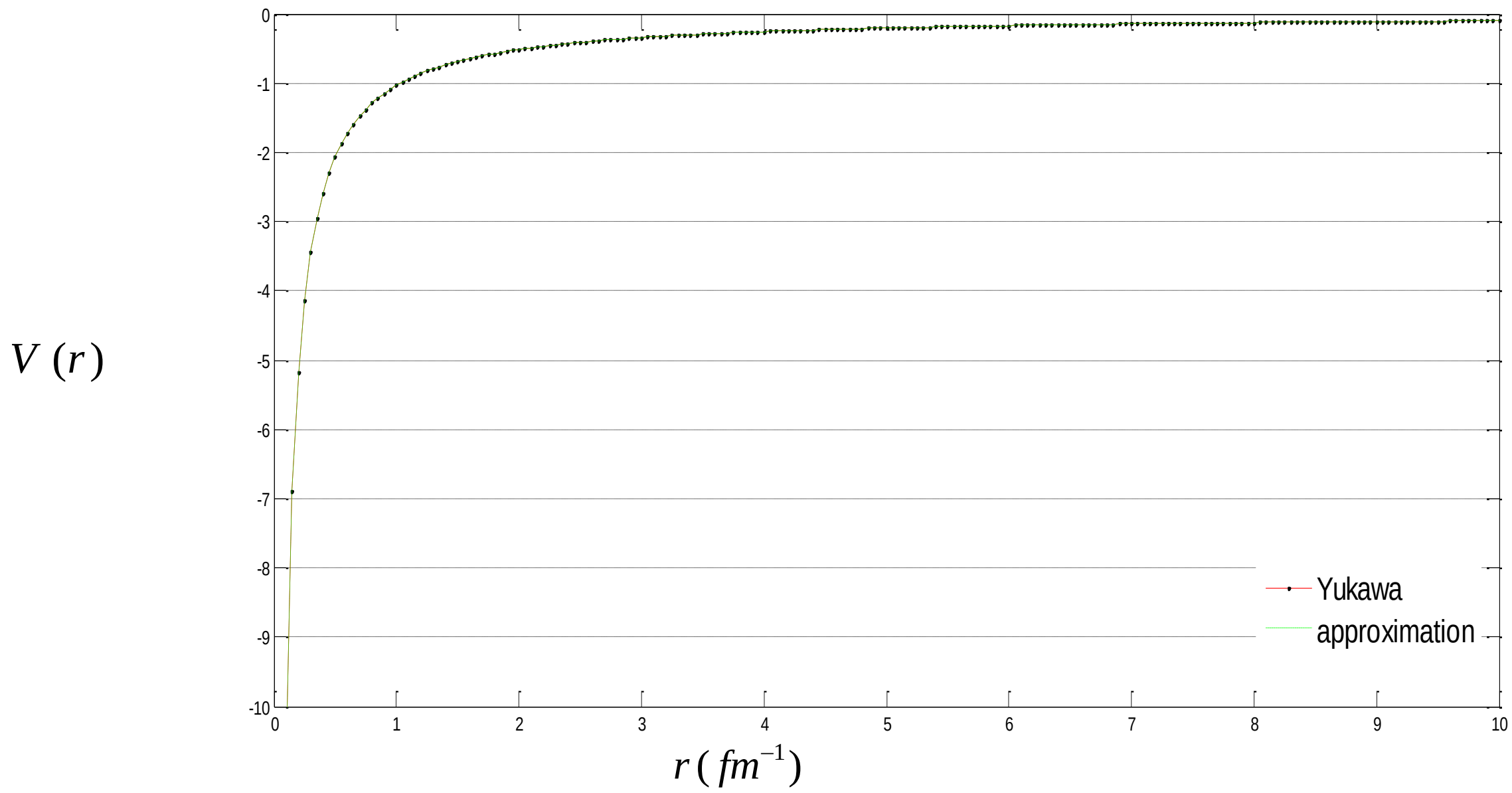

**Figure 1:** The Yukawa potential (red line) and its approximation in Eq. (16) (green line).